\documentstyle[editedvolume]{crckapb} 

\begin{opening}
\title{Ground Based VLBI Facilities -- the European and Global VLBI Network}

\author{M.A. Garrett} 
\institute{Joint Institute for VLBI in Europe\\
           Postbus 2, 7990 AA Dwingeloo, The Netherlands}

\end{opening}

\runningtitle{Ground-based VLBI Facilities: the EVN and Global VLBI}

\begin{document}



\section{Introduction}

There are at least four well known Very Long Baseline Interferometer
(VLBI) networks that provide open access to astronomers around the
world. These include:
\begin{itemize}
\item European VLBI Network, {\tt www.evlbi.org}
\item Very Long Baseline Array, {\tt www.aoc.nrao.edu/vlba/html/VLBA.html}
\item Coordinated Millimeter VLBI Array, {\tt web.haystack.mit.edu/cmva}
\item Australian Long Baseline Array, {\tt www.atnf.csiro.au/vlbi} \& the 
Asia-Pacific Telescope, {\tt www.atnf.csiro.au/apt}.
\end{itemize} 

For a comprehensive guide to the various array characteristics,
corresponding correlator capabilities, range of diverse observing
modes, different proposal submission procedures, user support {\it
  etc.},  I refer the reader to the on-line web pages highlighted above.  In
this paper I have chosen to focus on a sub-set of these facilities, in
particular the European VLBI Network (EVN) both in stand-alone mode but
also as a major component of a Global VLBI Network, involving MERLIN
(see {\tt www.merlin.ac.uk}) and the Very Long Baseline Array (VLBA --
see Zensus, Diamond \& Napier 1995, for a comprehensive review).

\section{The European VLBI Network} 

The EVN is a ``part-time'' VLBI network that observes in 3 ``block
sessions'' per year. Each of these block sessions is 3-4 weeks long, and
usually 3 or more different observing frequencies are available
within any given session. Observing sessions are scheduled in
February-March, May-June and October-November of each year and often
involve both Global and EVN-only observations. The EVN and Global VLBI
``Call for Proposals'' is issued three times per year with deadlines of
February 1, June 1, and October 1. Please refer to the web-based {\it
EVN User Guide} at {\tt www.evlbi.org} for more details on how to apply
for EVN observing time.

The locations of the telescopes, and in addition the EVN MkIV
Correlator at JIVE (Joint Institute for VLBI in Europe), are shown in
Figure 1.  Members of the EVN with radio telescopes are listed in table
1. There are several categories of EVN membership -- these recognise the
different levels of commitment by the various participating institutes
-- full or associate membership, and affiliated telescopes. These three
categories are separated in table 1 by dividing lines. The table also
indicates the diameters of the individual telescopes and their system
noise in Jy (at the main EVN observing wavelengths).

\begin{figure}
\vspace{7cm}  
\includegraphics{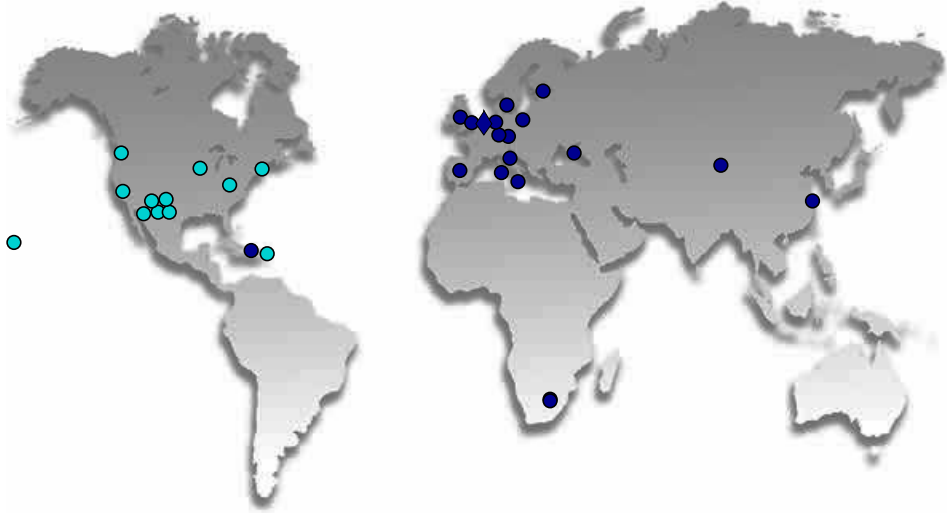}
\caption{The Global VLBI Network of radio telescopes, including the
  European VLBI Network (dark dots, including the new 64-m telescope In
  Sardinia) and the VLBA (lighter dots, including the GBT and
  VLA). The EVN and VLBA often co-observe forming a very sensitive,
  high resolution global VLBI network.}
\label{gvlbi} 
\end{figure}

\begin{table*} 
\begin{tabular}{|l|l|c|c|} 
\label{table1} 
Institute & Telescope & Diam (m) &  SEFD (Jy) \\
          &           &          &  $\lambda\lambda$21$|$18$|$6$|$5$|$4$|$1$|$0.7 cm     \\
\hline
MPIfR (DE) & Effelsberg & 100  & 20$|$19$|$20$|$55$|$20$|$140$|$600\\ 
ASTRON (NL) & WSRT & $14\times25$  & 30$|$30$|$60$|$--$|$120$|$--$|$--\\  
JBO (UK) & Lovell & 76 & 35$|$35$|$25$|$25$|$55$|$ - \\  
         &    Mk2 & 25 & 350$|$320$|$320$|$910$|$--$|$910$|$--\\  
         & Cambridge & 32 & 220$|$212$|$136$|$136$|$--$|$720$|$--\\ 
IRA (IT)   & Medicina   & 32  & 390$|$582$|$296$|$900$|$270$|$1090$|$2800 \\  
           & Noto       & 32  &  820$|$784$|$260$|$--$|$770$|$2500$|$3000\\ 
           & Sardinia   & 64  &  Not yet known \\ 
OSO (SE)   & Onsala-85  & 25  &  450$|$390$|$600$|$1500$|$--$|$--$|$--\\   
           & Onsala-60  & 20  &  --$|$--$|$--$|$--$|$1630$|$1380$|$1310\\  
SHAO (CN)  & Shanghai   & 25  &  --$|$1090$|$520$|$--$|$590$|$1606$|$--\\ 
UAO (CN)   & Urumqi     & 25  &  1068$|$1068$|$353$|$--$|$396$|$2950$|$--\\ 
TCfA (PL)  & Torun      & 32  &  250$|$230$|$250$|$300$|$--$|$--$|$--\\ 
OAN (ES)   & Yebes 14-m & 14  &  --$|$--$|$--$|$--$|$3300$|$--$|$4160\\ 
           & Yebes 40-m & 40  &  Not yet known \\ 
\hline 
MRO (FI)   & Metsahovi  & 14  &  --$|$--$|$--$|$--$|$--$|$2608$|$4500 \\ 
NAIC (USA) & Arecibo    & 305 &  3$|$4$|$6$|$9$|$--$|$--$|$-- \\ 
HRAO (ZA)  & Hartebeesthoek & 26 & --$|$450$|$700$|$800$|$940$|$--$|$-- \\        
IfAG (DE)  & Wettzell       & 20 & --$|$--$|$--$|$--$|$750$|$--$|$-- \\ 
\hline          
DSN (USA$|$ES) & Robledo 70-m & 70  & --$|$42$|$--$|$--$|$23$|$100$|$--\\  
             & Robledo  34-m & 34  & --$|$--$|$--$|$--$|$88$|$--$|$-- \\ 
CGS$|$ASI (IT) & Matera   & 20  & --$|$--$|$--$|$--$|$900$|$--$|$-- \\               
NMA (NO) & Ny-Alesund   & 20  & --$|$--$|$--$|$--$|$1255$|$--$|$-- \\         
CrAO (UA)  & Simeiz         & 22 & --$|$1600$|$--$|$3000$|$--$|$1200$|$3000$|$--\\ 
\hline
\end{tabular} 
\caption{EVN member telescopes including those currently under
          construction at Sardinia and Yebes. System Equivalent Flux
          Density (SEFD) values for new or refurbished telescopes/receivers
          (e.g. the re-surfaced Lovell 76-m) are current best estimates
          or as yet unknown. 
          The table is divided into
          3 parts: full, associate members and affiliated observatories.}

\end{table*} 

\subsection{EVN Sensitivity} 

The obvious advantage that the EVN has over other networks is the
enormous collecting area it can routinely draw upon across a broad
range of frequencies. It is worth remembering that in terms of
collecting area, the larger EVN telescopes such as the Effelsberg
100-m, Westerbork tied array and Lovell 76-m, are either individually
larger or comparable to the combined collecting area of the ten 25-m
antennas that comprise the VLBA. In practical terms, this large
collecting area permits much fainter sources to be detected, imaged and
self-calibrated with the EVN. This advantage applies particularly to
spectral line observations (both emission and absorption studies). In
addition, another key advantage of the EVN is its capability to perform
{\it sustained} VLBI observations at very high data rates (currently
512 Mbit/sec). Uniquely, the EVN is able to observe at these sustained
data rates for 12 hours or longer.

Figure 2 shows the $1\sigma$ r.m.s.  (image) noise level of a standard
EVN array (excluding the DSN telescopes and Arecibo) at $\lambda18$~cm
for typical spectral line single channel widths of 6 and 30 kHz, dual
polarisation continuum bands of 16~MHz (128 Mbits/sec), 32~MHz (256
Mbits/sec) and 64~MHz (512 Mbits/sec). As a point of reference I plot
the corresponding noise levels of the VLBA at 6~kHz, and 16 MHz (the
latter being equivalent to the VLBA's maximum sustainable data rate of
128 Mbits/sec over 12 hours). In the same figure I also include an
extremely sensitive Global VLBI array including the EVN, VLBA, VLA,
GBT, DSN and Arecibo. Casting an eye towards the future, I have plotted
the noise level that would be achieved by the EVN at $\lambda 6$~cm,
assuming fibre connections ($e$EVN) and a bandwidth of $\sim 2$~GHz per
polarisation (see section \ref{future}). In the latter case, the $e$EVN
can be expected to reach impressive sub-microJy noise levels in a
typical 12 hour (on-source) observing run.

\begin{figure}
\vspace{8cm}  
\includegraphics{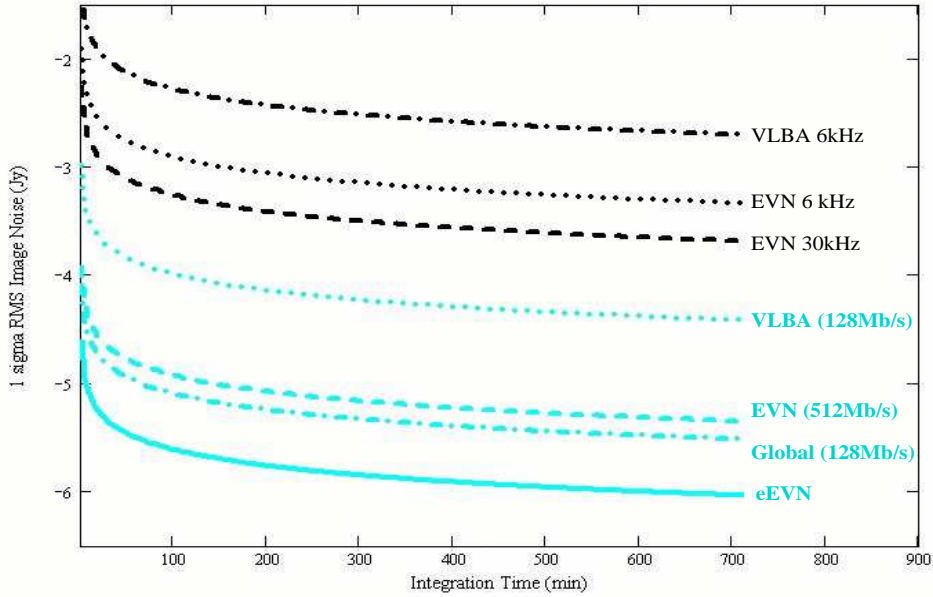}
\caption{The 1-$\sigma$ image noise level achieved by the EVN at
  $\lambda 18$~cm 
  as a function of time (minutes) for typical spectral line channel
  widths (6 and 30~kHz), dual-polarisation continuum data rates of 128,
  256 and 512 Mbit/sec. For comparison the noise levels achieved by the VLBA
  for a channel width of 30~kHz and its maximum sustained data rate of
  128 Mbits/sec are also included. The noise level achieved by a global
  VLBI array operating at a sustained data rate of 128 Mbits/sec is
  also presented. Finally we present the anticipated performance of the $e$EVN
  -- the current EVN telescopes connected together by optical fibres.}
\label{gvlbi} 
\end{figure}

\subsection{Unique EVN frequencies} 

Another important feature of the EVN is the availability of observing
frequencies that are essentially unique, at least in the northern
hemisphere. These include UHF band  observations ($\sim 800-1300$~MHz)
that have been used to search in relatively distant extra-galactic
systems for (redshifted) neutral hydrogen in absorption, and the
$\lambda 5$~cm receivers that have been used to infer the presence of
circumstellar discs around massive stars located within star forming
regions in our own galaxy (see Booth these proceedings for a summary of
the latest extra-galactic HI absorption and Methanol maser
results). The fact that these receivers were constructed quickly, and
then rapidly deployed across a substantial fraction of the network,
emphasises the EVN's ability to respond flexibly to ``bottom up'',
user driven demand.

\subsection{Combined joint EVN-MERLIN observations} 

The EVN often co-observes with the UK's MERLIN radio telescope network.
Two of the MERLIN telescopes conduct both VLBI and MERLIN observations
simultaneously -- usually the Cambridge 32-m telescope and one of the
Jodrell Bank ``home'' telescopes (either the Lovell 76-m or Mk2 telescope).
The advantage of joint EVN-MERLIN observations is the excellent
uv-coverage that can be obtained from the combined data set. MERLIN
provides baselines on scales ranging from 6 to 217~km, thus providing
overlap with the shortest (projected) EVN baselines (Jodrell-Cambridge
and Effelsberg-Westerbork in particular). The combined data set
therefore includes a range of baseline lengths, from a few to several
thousand kilometers. This is of course ideal for detecting and
imaging large extended sources that might otherwise be resolved-out or
be extremely difficult to  image accurately with the EVN alone.

Usually EVN-MERLIN observations take place during every session at one
of the main EVN observing frequencies (usually $\lambda\lambda18$ or
6~cm, although joint observations at $\lambda$1~cm are also possible).  The
inclusion of the common Jodrell-Cambridge baseline in both the EVN and
MERLIN arrays, ensures that the data sets can be combined together in a
consistent fashion.

\subsection{The EVN as part of the Global VLBI Network}

VLBI is an international effort. The very longest baselines available
via the ground require collaborations between various VLBI networks. In
the northern hemisphere a particularly strong collaboration exists
between the EVN and the VLBA. Both networks employ very similar
recording systems (MkIV and VLBA respectively) which provide a wide 
range of compatible observing modes. Global VLBI observations usually
involve the participation of the most sensitive VLBI telescopes in the
world, including those that provide the longest baseline lengths -- it's
not uncommon for up to 20 VLBI telescopes to participate in a single 12
hour observing run.  In addition, many Global VLBI projects are also
made together with simultaneous MERLIN observations. This high
resolution, Global VLBI Network currently provides the ultimate in
terms of both sensitivity and uv-coverage (see Figure 3). Snapshot
observations of a large number of sources, or of galactic sources that
evolve quickly also becomes feasible with such a 20 station global array.

\subsection{On-going enhancements to the EVN} 

So far we have focussed on the areas where the EVN is strongest, {\it
  viz.} unmatched sensitivity and the ability to observe at several
unique frequencies. There is a continuous and vigorous EVN programme of
development to maintain and enhance these capabilities. General
enhancements to the network (e.g. the recent upgrade to 2-head
recording and 512 Mbit/sec data rates) are coordinated via the EVN
Technical \& Operations Group (see {\tt www.evlbi.org/tog/tog.html} for
more details).

Several significant events are expected to take place in the
short-term, in particular the addition of two large telescopes to the
EVN over the course of the next two to three years.  These are the 64-m
Sardinia Radio Telescope (the SRT -- to be built at San Basilio, near
Cagliari and operated by the IRA) and the OAN-Yebes 40-m telescope (to
be built alongside the current 14-m OAN antenna at Yebes, near Madrid).
In 2005 the Miyun 50-m mesh telescope (located near Beijing, China)
should be complete, and this may also participate in EVN observations
(up to and including $\lambda3.6$~cm). In addition, several other major
upgrades of EVN telescopes have just been completed. These include the
recent upgrade of the WSRT array, and the installation of an active
surface for the Noto 32-m telescope. The replacement of the existing
reflecting surface of the Lovell 76-m telescope is on-going and is
expected to be complete by the end of 2002. These will permit the
Lovell telescope to observe usefully at frequencies up to 10~GHz --
boosting its sensitivity by a factor of 5 at $\lambda 6$~cm. This major
engineering development (which includes the upgrade of the drive and
pointing control system) will transform the Lovell's capability as a
VLBI and MERLIN antenna.  Progress with the Lovell telescope upgrade
and the construction of the new 40-m telescope at Yebes (as of summer
2001) is presented in Figure 4.

It must also be noted that there are certainly areas where the
performance of a heterogeneous network such as the EVN might be
considered less than optimal, at least in comparison to a homogeneous,
full-time network such as the VLBA.  Certainly the EVN is a more
difficult instrument to calibrate, and only recently have a significant
number of telescopes achieved frequency flexibility. In addition, the
geographical location of the majority of the antennas is also not
optimal for high frequency observations ($> 20$~GHz). The EVN's ability
to react to ``target-of-opportunity'' observations is also more limited
than the VLBA -- at least outside of network sessions.  Similarly it is
difficult for the EVN to adequately monitor sources with evolving radio
structure, at least in comparison to the uniform temporal coverage that
the VLBA can provide.

Nevertheless, progress is being made in all these areas.  Automatic
pipelining of EVN (and global) VLBI data (see section \ref{user} and
Reynolds, Paragi \& Garrett 2002) now largely hides the intricacies of
EVN calibration from the user. Experiments requiring fast frequency
switching are beginning to become more common in network sessions, and
the addition of the new 40-m Yebes and 64-m Sardinia telescopes
(capable of operating at frequencies up to 115 GHz) will enhance the
EVN's sensitivity at higher frequencies. Vigorous efforts to move the
EVN towards real-time operations (see section \ref{future}) will also
provide increased flexibility to conduct more uniform monitoring
campaigns or to respond to ``target-of-opportunity'' events.

\begin{figure}[h]
\vspace{3.8cm}  
\includegraphics{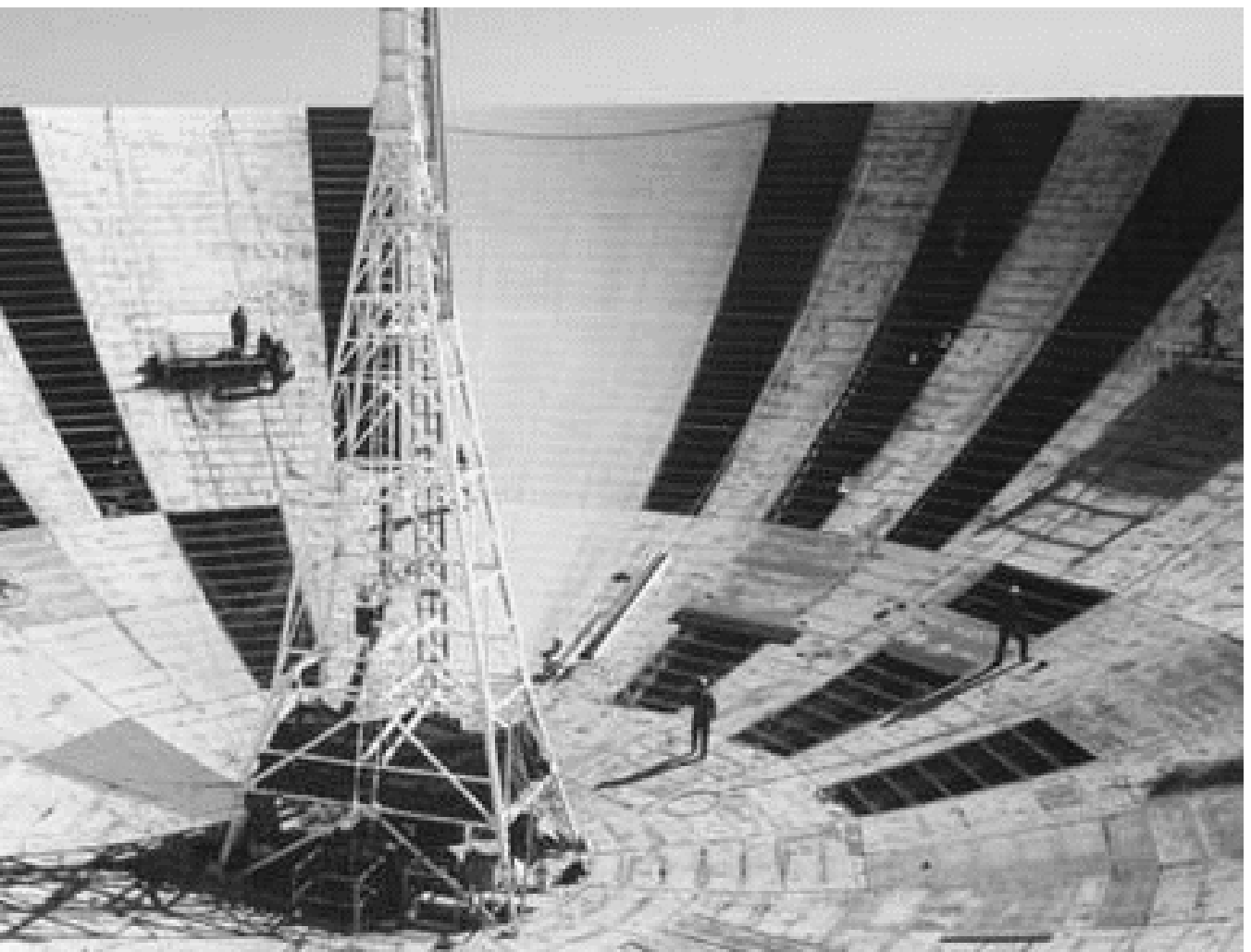}
\includegraphics{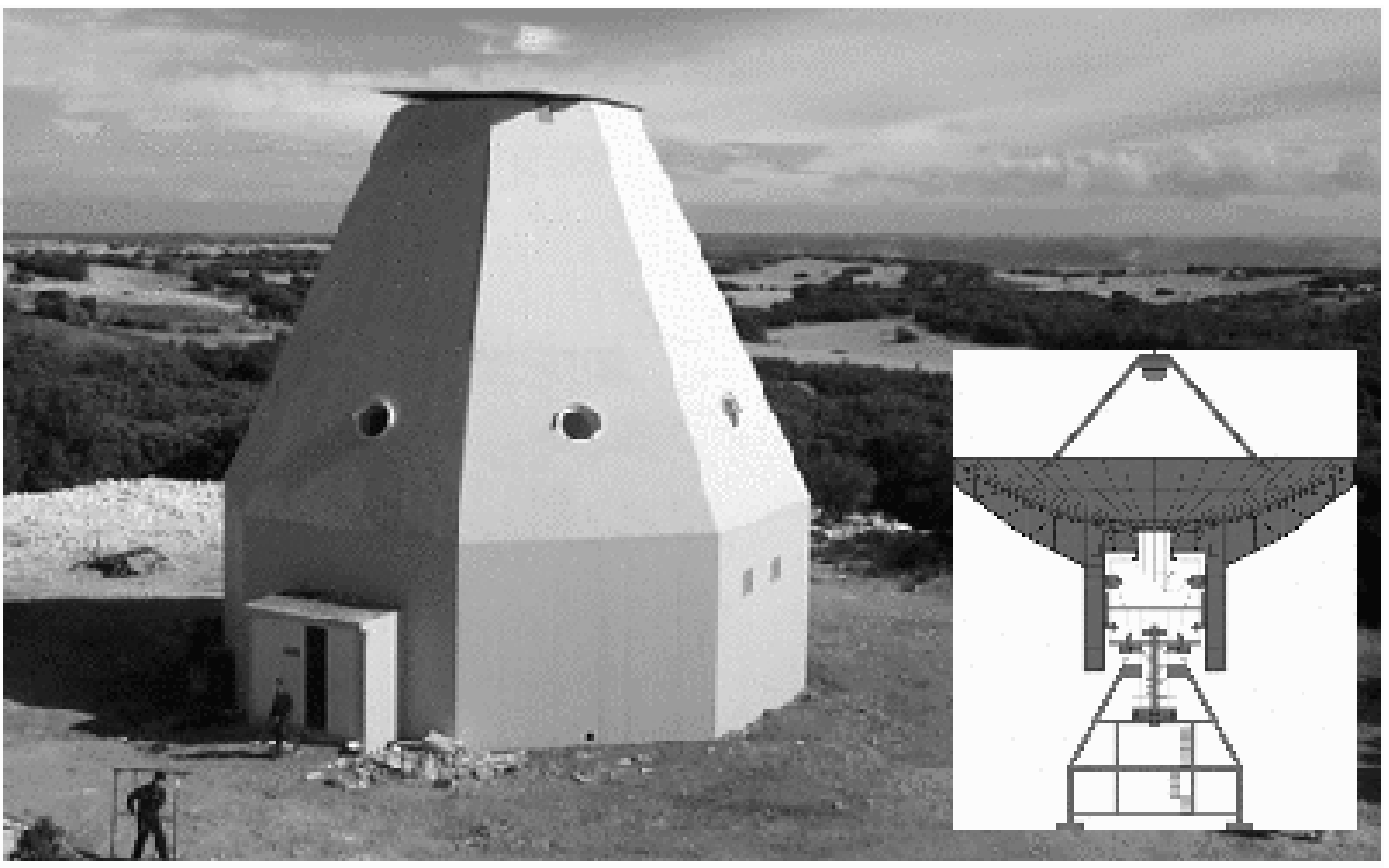}
\caption{Left: The upgrade of the Lovell Telescope surface -- a factor
  of 5 improvement in sensitivity is expected at $\lambda6$~cm. Right:
  Construction continues of the 40-m telescope at Yebes -- the pedestal
  is complete, work continues on the backing structure and reflector
  panels. The new 40-m telescope (inset) will operate at frequencies up to
  115~GHz}
\label{fir-rc} 
\end{figure}

\section{The EVN MkIV Data Processor at JIVE} 

The construction and development of a VLBI correlator entirely
dedicated to EVN activities is one of the great achievements of the
last decade. The EVN MkIV Data Processor at JIVE (Casse 1999, Schilizzi
et al. 2002b) was developed as part of an international collaboration,
the primary contributors being the European Consortium for VLBI
and the MIT Haystack Observatory, in the USA.  The EVN MkIV Data
Processor is operated by JIVE and is now the main-stay of EVN data
correlation (including global projects which it shares with the
NRAO-VLBA).

The Data Processor is capable of handling data from 16 telescopes
simultaneously (more via multiple pass correlation) and can handle
MkIV, VLBA and MkIII data formats.  Standard correlation of the vast
majority of EVN and global VLBI continuum and spectral line experiments
are now routinely processed at JIVE. The capacity of the correlator is
continually being enhanced, and new capabilities introduced (see {\tt
  www.jive.nl} for the most up-to-date information).  On-going projects
include: recirculation (in order to provide superb spectral resolution
in excess of 8192 channels per baseband), Pulsar Gating (to optimise
Pulsar detection limits) and the PCInt (Post-correlator Integrator)
that will permit high-speed read-out of the correlator at data rates of
up to 160 MBytes/second). 

The PCInt development is expected to see ``first-light'' by the end of
2002 (Parsley 2001a) -- it will transform the capability of the EVN (and
Global VLBI arrays) providing the possibility to image dozens of faint
sub-mJy radio sources within the primary beam of the individual VLBI
antennas (see Garrett these proceedings). To take advantage of the
fantastic output data rates the PCInt can generate, significant
off-line computing resources will be required. Off-line computing
resources are likely be the main bottleneck in the new system, at least
for the first few years of operation.

\subsection{EVN User Support} 
\label{user} 

As well as operating the EVN Data Processor, JIVE is largely
responsible for EVN user support. JIVE support scientists (and in
addition other JIVE staff) are involved in providing a level of user
support that was previously unknown within the EVN, and rivals or
surpasses that provided by other instruments. In particular, the 
following services are routinely provided:

\begin{itemize} 
\item advice regarding the technical content of proposals
  (e.g. cover-sheet specifications, choice of mode, observing strategy etc)
\item scheduling assistance and maintenance/development of NRAO's Sched (for
  specific EVN requirements)
\item absentee correlation and data quality check-out 
\item automatic calibration of EVN and Global VLBI data correlated at
  JIVE (via a Pipeline process) 
\item direct assistance with VLBI data and image analysis.  
\end{itemize} 

In addition, the support scientists also contribute to monitoring the
reliability and performance of the EVN (via special Network Monitoring
Experiments) and also conduct network tests aimed at extending the
capabilities of the network.

Financial support is available to those EVN users that wish to visit
JIVE in order to avail themselves of these services (in particular
scheduling and data analysis). Indeed the EVN is in receipt of a
substantial award from the European Commission in Brussels ({\it Access
  to Research Infrastructures}), that comprehensively supports EVN users
that are not affiliated to the EVN Consortium institutes but are
located within the European Union or Associated States. In addition,
there is also internal EVN support for users that are directly
affiliated to EVN Consortium institutes. Both programmes support not
only visits to JIVE but also to other members of the EVN. For example,
users frequently visit Jodrell Bank Observatory in order to 
take advantage of the local expertise in combining joint
EVN-MERLIN data sets. 

\begin{figure}
\vspace{5.7cm}  
\includegraphics{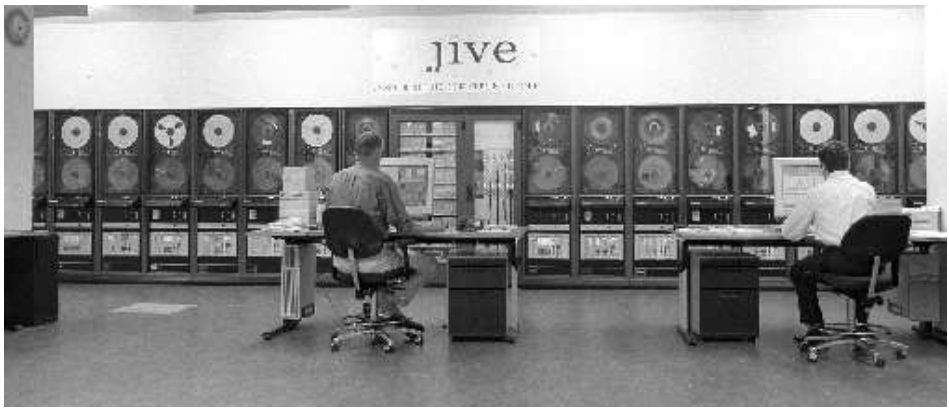}
\caption{The EVN MkIV Data Processor at JIVE -- the 16 station Data
  Processor is now the main-stay of EVN data correlation.}
\end{figure}

\section{The Future of the EVN}  
\label{future} 

Across the globe radio telescopes and interferometer arrays are
involved in significant efforts to improve their overall performance
and their continuum sensitivity in particular. These developments are
desperately needed in order for radio astronomy to maintain its
competitiveness with other next generation instruments -- especially
sub-mm and IR telescopes. Anticipated improvements largely rely on the
possibility of observing and processing much larger continuum
bandwidths than was previously possible. The use of optical fibre
technology now permits the digital transport of many GHz of bandwidth
over 1000's of km. Next generation correlators are now being designed
and constructed to handle the associated Gbit/sec input data rates and
subsequent processing requirements.  The EVLA, e-MERLIN and LOFAR
telescopes will be the first radio telescopes to take advantage of
these developments -- permitting huge areas of sky to be mapped-out
with sub-arcsecond resolution and microJy sensitivity.

\begin{figure}[h]
\vspace{6.0cm}  
\includegraphics{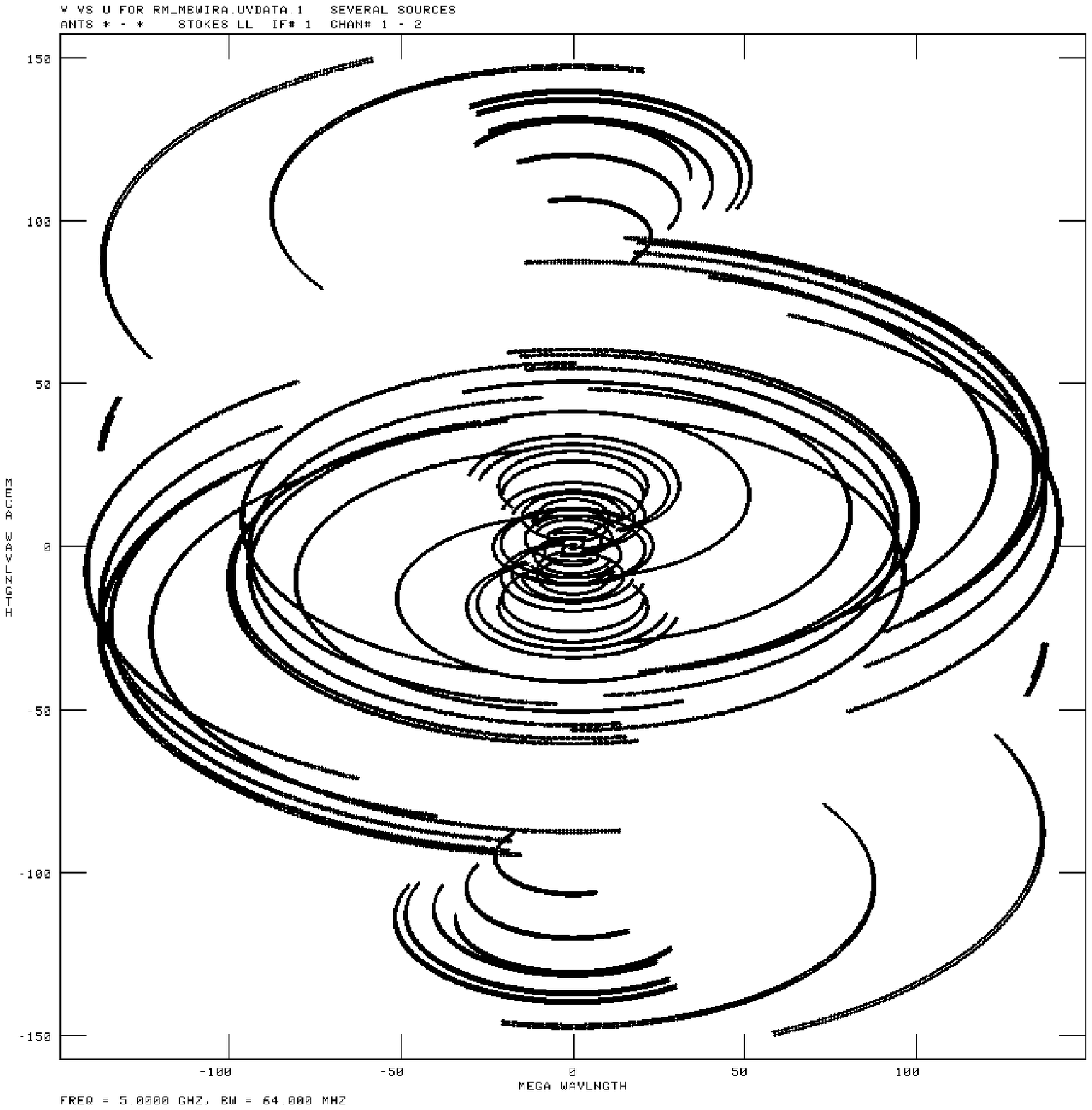}
\includegraphics{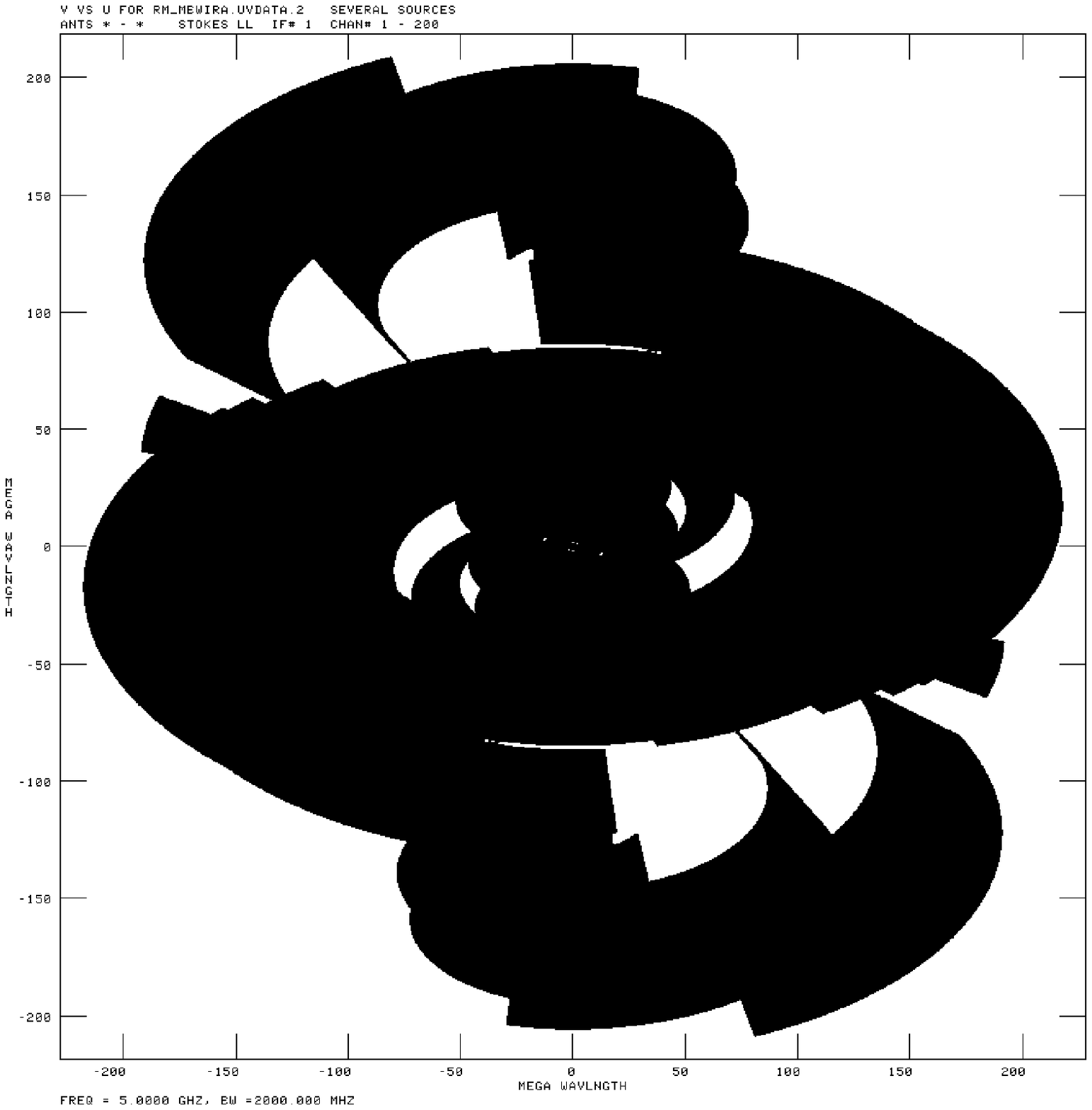}
\caption{Left: The current uv-coverage of the EVN at $\lambda6$~cm for a source
  located at $\delta=30^{\circ}$. Right: the extended (almost full)
  uv-coverage of the $e$EVN for the same source, assuming a total
  bandwidth of 2~GHz.}
\label{uvcov} 
\end{figure}

The consequences for VLBI, and the EVN in particular, are crystal
clear. In order to remain competitive with, and complementary to these
upgraded or new radio instruments, the EVN {\it must} be able to
observe and process several GHz of bandwidth too. The connection of the
EVN telescopes to commercial ``$\lambda$-networks'' (fibre data
transport utilising wavelength-division multiplexing techniques) is now
being vigorously explored in both Europe and the USA - the first
connections and fringe tests are expected to occur within the year
(Schilizzi 2002a, Whitney 2002, Parsley 2002).  Perhaps all VLBI
telescopes (in particular those located nearby densely populated areas)
will be connected to such networks in 5-10 years time, the exact
time-scale depending on local circumstances. Trans-continental
connections also appear feasible too.  Meanwhile the new disk-based MkV
(Whitney 2001) and PC-EVN (Parsley 2001b) recording systems are set to
replace the current generation of tape recording systems. These
disk-based PC systems can already record data at rates that are similar
to current MkIV or VLBA systems.  In addition, the same systems are
poised to take advantage of the expected expansion in the capability of
PC hardware over the next few years.  While the investment in both
fibre or disk-based technologies is substantial, it will provide VLBI
networks such as the EVN with sensitivity levels that are similar or
even better than that anticipated for either the EVLA or e-MERLIN. A
natural consequence of employing observing bandwidths that span several
GHz is the almost complete uv-coverage that accompanies it. Figure
\ref{uvcov} shows the uv-coverage of a fibre connected EVN ($e$EVN).
The transparent, real-time combination of the $e$EVN and e-MERLIN will
also result in a significant enhancement in imaging capabilities of the
combined array.

As a consequence of all these developments, a replacement for the EVN
Data Processor will also be necessary. The new correlator will need to
be capable of handling a global array of $\sim 30$ telescopes (each
generating 10-30 GBits/sec of data) and the phenomenal output data
rates that will enable the natural field of view (then set by the
primary beam of individual VLBI elements) to be imaged out in its
entirety. New broad-band receivers and a new generation of VLBI data
acquisition electronics will also be required at the telescopes, in
order to take full advantage of the available bandwidth. There are (not
surprisingly) severe implications for (``off-line'') data processing
requirements too. This not only concerns raw processing power but also
the development of new calibration and image algorithms. In a very real
sense, the $e$EVN with baselines on the scales of several thousand km,
``fantastic'' data rates and microJy sensitivity, will be the natural
test-bed to investigate some of the problems and possible limitations
that might be relevant to next generation instruments such as the SKA
(see Kus these proceedings).

\end{document}